\documentclass[prd,twocolumn,unsortedaddress,superscriptaddress]{revtex4-2}
\usepackage[english]{babel}
\usepackage{graphicx}
\usepackage{epsfig,color}
\usepackage{amsmath,amssymb,eufrak,calc}
\usepackage{amsthm}
\usepackage{enumerate}
\usepackage{hyperref}
\usepackage{amsxtra,amsfonts,bm}
\usepackage{blkarray}
\usepackage{adjustbox}

\begin{document}

\title{Duality as a Feasible Physical Transformation}

\date{\today}

\author{Shachar Ashkenazi}
\address{Racah Institute of Physics, The Hebrew University of Jerusalem, Jerusalem 91904, Givat Ram, Israel.}

\author{Erez Zohar}
\address{Racah Institute of Physics, The Hebrew University of Jerusalem, Jerusalem 91904, Givat Ram, Israel.}

\begin{abstract}
Duality transformations are very important in both classical and quantum physics.  They allow one to relate two seemingly different formulations of the same physical realm through clever mathematical manipulations, and offer numerous advantages for the study of many-body physics. In this work, we suggest a method which shall introduce them to the world of quantum simulation too: a feasible scheme for  implementing duality transformations as physical operations, mapping between dual quantum states showing the same observable physics,  rather than just a mathematical trick. Demonstrating with Abelian lattice models, we show how duality transformations could be implemented in the laboratory as sequences of single- and two-body operations - unitaries and measurements. 
\end{abstract}

\maketitle

\tableofcontents

\section{Introduction}

Duality is a very deep and meaningful concept, spanning across various scientific disciplines. In physics, duality between two models implies that they are both valid mathematical descriptions of the same observable physics, either classical of quantum.

Dual models may differ from one another in many ways. In the case of self-dual models, the two ends of the duality transformation are similar mathematically, with the same Hilbert spaces, degrees of freedom and symmetries, perhaps only defined on shifted positions (as in the case of the Ising and clock models in a single space dimension \cite{fradkin_order_1978,savit_duality_1980}). Taking the same models to higher dimensions serves as a good example to duality transformations between models with different Hilbert spaces and symmetries: clock models in $2+1d$, which are lattice models whose degrees of freedom reside on the sites, manifesting a global symmetry, are dual to $\mathbb{Z}_N$ lattice gauge theories \cite{fradkin_order_1978,elitzur_phase_1979,horn_hamiltonian_1979}, where the degrees of freedom reside on the links, but the symmetry is local. While in this example, the dual models are still defined on similar manifolds and geometries, this is also not a necessary condition for the duality of physical models: consider, for example, the AdS-CFT correspondence, linking gravitational models in anti-deSitter spaces to conformal field theories residing on the AdS boundary \cite{maldacena_the_1999}. 

Duality transformations  provide some exact correspondence between possibly different physical theories, which is a beautiful, fundamental concept on its own.  Moreover, the study of a physical model can be made, in some cases, significantly simpler by switching to its dual formulation. For example, even in the simple case of self dual models as in the $1+1d$ Ising case, in spite of the fact that the dual model is the same, strong couplings are mapped to weak ones, which has many advantages \cite{fradkin_phase_1979}. Duality transformation, as argued above, can map a model with a local symmetry to one with only a global one, hence systematically reducing  the number of constraints in the system (from an extensive number to a minimal, global set of constraints), the redundancy of the Hilbert space (and hence its size). This has been very useful in several studies of lattice gauge theories \cite{fradkin_order_1978,drell_quantum_1979,elitzur_phase_1979,horn_hamiltonian_1979,kaplan_gauss_2018,bender_gauge_2020,bender_real_2020,haase_resource_2021,paulson_simulating_2021}.

Furthermore, quantum simulation \cite{Cirac_goals_2012,georgescu_quantum_2014} - in which one maps a hard to solve quantum model to another one which can be implemented, manipulated and measured in the laboratory (following Feynman's suggestion \cite{feynman_simulating_1982}) - is a valid computational method in physics these days, allowing us to deal with a variety of nonperturbative many-body problems. While there are several approaches to the design of quantum simulators, they quite often require the implementation of complicated interactions that are not very natural for the simulating platforms. Duality can be beneficial here too: when simulating quantum models with a duality transformation, one may find that in different cases, different sides of the duality transformations are easier to implement. By possibly allowing to weaken the coupling strength, convert interacting terms to non interacting ones (or reduce the number of interacting bodies), simplify (or completely eliminate) constraints to which a physical system is subject, duality transformations can truly simplify quantum simulation.

Normally (and in most cases), duality transformations are pure mathematical manipulations, which remain completely theoretical, in spite of their important physical meanings. However, when duality transformation is used in a quantum simulator, it could be useful to allow it to describe the system in both dual forms (both sides of the duality transformation) rather than simulating separately different physical regimes, using different simulators. Including both ends of the duality transformation in the same simulator requires a larger Hilbert space -  a product of the Hilbert spaces hosting the degrees of freedom of both dual formulations; and, obviously, one should be able to implement the duality transformation as a \emph{physical operation}; such an operation is expected to take a state in one formulation of the theory, which we call $A$, put it in a product state with some idle state in the dual space $B$, which exists in parallel, and convert it to a product of an idle state in $A$ and a dual state in $B$, containing the same physical information, but in terms of the dual degrees of freedom. 

Moreover, if duality transformations are experimentally feasible, this may not only be used as a tool for the quantum simulation of models admitting duality transformations; it can also be useful, in some cases, to explicitly carry out a duality transformation, experimentally, and observe its consequences, that is, making the duality transformation the subject of the quantum simulation, rather than a tool.

If we want a duality transformation to be physically feasible, it needs to be composed out of unitary operations and measurements at most; feasibility will be increased if it can be factorized into simple operations, e.g. single- and two-body terms. In this work, we will demonstrate that this is indeed possible for a variety of physical models, focusing on lattice models with Abelian symmetries, by explicitly constructing such physical duality transformations, that will be invertible and, in some sense, feasible.

We will begin by briefly explaining the main ideas behind the procedure we propose. Then, we will demonstrate it in detail for $1+1d$ clock models, extend the discussion to higher dimensions by constructing the $2+1d$ example, and finally generalize to the $U(1)$ case.

\section{Background and Concept}
In this work we consider two separate physical systems, denoted by $A$ and $B$, with Hilbert spaces $\mathcal{H}_A$ and $\mathcal{H}_B$ respectively.
We assume that the Hamiltonian $H$ of system $A$ is invariant under some global symmetry:
let $G$ be some symmetry group; for each element $g \in G$, 
there exists a unitary transformation $\mathcal{Z}_g$ such that 
\begin{equation}
    \mathcal{Z}_g H \mathcal{Z}^{\dagger}_g = H.
\end{equation}
Therefore, there exists a maximal set of mutually commuting operators $\left\{\mathcal{Z}_g\right\}$ which can be diagonalized simultaneously with the Hamiltonian $H$. Their mutual eigenstates will define global \emph{selection sectors}: separate subspaces of $\mathcal{H}_A$ which cannot be connected by the dynamics governed by $H$. In this work we consider Abelian symmetries, where the symmetry group is either $\mathbb{Z}_N$ or $U(1)$, and thus these global sectors are determined by the eigenstates of a single transformation operator $\mathcal{Z}$ (or a single symmetry generator in the $U(1)$ case), and labelled by the respective $\mathcal{Z}$ eigenvalue.

Let $\mathcal{O}\in O\left(\mathcal{H}_A\right)$ be an observable operator acting  on $\mathcal{H}_A$. It will be physically relevant only if it is invariant under the symmetry action, that is, $\mathcal{ZOZ}^{\dagger}=\mathcal{O}$; otherwise, its expectation value will always vanish (disregarding spontaneous symmetry breaking \cite{englert_broken_1964,higgs_broken_1964}, which we do not discuss in this work).

In this work we map, using a sequence of local unitary operations and measurements, sectors of the system $\mathcal{H}_A$ to sectors of a \emph{dual} system $\mathcal{H}_B$, in a special procedure we introduce. In the dual system, the physics is described by the Hamiltonian $\tilde{H}$, invariant under a global symmetry operation $\tilde{\mathcal{Z}}$. We will show how to relate the Hamiltonians, the states and the observables: each  $\left|\psi\right\rangle_A$ invariant under $\mathcal{Z}$ in a sector of $\mathcal{H}_A$ will be mapped to a $\left|\tilde{\psi}\right\rangle_B$ invariant under $\tilde{\mathcal{Z}}$ in an appropriate sector of $\mathcal{H}_B$; similarly, invariant operators $\mathcal{O}\in\mathcal{H}_A$, for which $\mathcal{ZOZ}^{\dagger}=\mathcal{O}$ (and thus are block diagonal in the selection sectors), will be mapped to invariant operators 
$\tilde{\mathcal{O}}\in\mathcal{H}_B$ for which $\tilde{\mathcal{Z}}\tilde{\mathcal{O}}\tilde{\mathcal{Z}}^{\dagger}=\tilde{\mathcal{O}}$, block-diagonal in the global sectors of $\mathcal{H}_B$.

Thanks to the locality and simplicity of our procedure, it allows one to implement the duality transformation, which usually is merely a mathematical map, as a real, \emph{ physical operation}, which can be carried out experimentally, between two actual Hilbert spaces $\mathcal{H}_A,\mathcal{H}_B$. This can be useful in the context of quantum simulation \cite{Cirac_goals_2012} either as a simulation resource or the simulation's target; that is, either for studying different coupling regimes of some model using a quantum simulator, or for the purpose of physically implementing a duality transformation.

\section{Our Procedure}

We say that two physical models, $A$ and $B$, are dual to one another, if there is a  way of mapping 
systems $A$ and $B$ which leaves the physics invariant. Each state $\left|\psi_i\right\rangle \in \mathcal{H}_A$ is mapped to a state $\left|\tilde{\psi}_i\right\rangle \in \mathcal{H}_B$, and each observable $\tilde{\mathcal{O}} \in O\left(\mathcal{H}_A\right)$ (that is, the set of operators acting on $\mathcal{H}_A$) is mapped to an observable  $\mathcal{O} \in \tilde{O}\left(\mathcal{H}_B\right)$ such that the \emph{matrix elements} are preserved,
\begin{equation}\label{dualitydef}
   \left<\psi_i\right|\mathcal{O}\left|\psi_j\right>=\left<\tilde{\psi}_i\right|\tilde{\mathcal{O}}\left|\tilde{\psi}_j\right>
\end{equation}.

The general idea of our map goes as follows.
Let system $A$ be in some state $\left|\psi\right\rangle\in\mathcal{H}_A$, which, thanks to the global symmetry, belongs to one of the sectors. 
We first embed it in the wider Hilbert space $\mathcal{H}_A \times \mathcal{H}_B$, in the form of the product state $\left|\psi\right\rangle_A \otimes \left|\tilde{\text{in}}\right\rangle_B$, where $\left|\tilde{\text{in}}\right\rangle_B$ is our choice of an \emph{initial} state of the dual system, to be introduced.
Next, using a speical local unitary transformation, we entangle the systems, which, after an appropriate post-selection of the state of $A$, will collapse to the form $\left|\text{out}\right\rangle_A \otimes \left|\tilde{\psi}\right\rangle_B $ - a product of a trivial, well known "idle" state of the original system $A$ with a state of the dual system containing exactly the same physical information of the original $\left|\psi\right\rangle_A$, hence implementing a duality transformation. 

\begin{figure}
    \centering
    \includegraphics[width=0.4\textwidth]{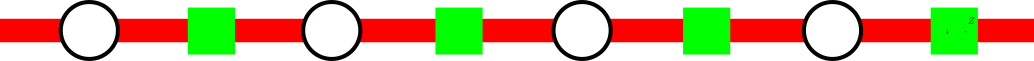}
    \caption{A one dimensional lattice: the black circles are sites of one system ($A$). The dual sites ($B$) are in between them, on the links (green squares). This is similar to the placement of matter and gauge fields in lattice gauge theories, on the sites and links respectively; however, when no further information is provided, one can see $B$ as the gauge fields of the matter $A$, or vice versa.}
    \label{fig:1d gauged}
\end{figure}

The degrees of freedom of the dual system $B$ reside on the lattice dual to this of $A$. Consider, for example, the case of a single space dimension: there, the local degrees of freedom of $B$ will reside on the links of $A$, and this is exactly where gauge fields reside in lattice gauge theories (See Fig. \ref{fig:1d gauged}) \cite{wilson_confinement_1974,kogut_hamiltonian_1975}. Gauge theories, thanks to the local symmetries upon which they are built,  involve local constraints (or Gauss laws). We will use these two facts and connect the systems in a minimal coupling procedure: first, we shall introduce the dual Hilbert space of the dual system, $\mathcal{H}_B$, as a space of gauge fields, making the global symmetry of $A$ local. On the states of the lattice gauge theory obtained that way, we will act with a unitary which flips the roles of gauge fields (now $A$) and matter (now $B$); finally, by performing the right measurement of the gauge fields $A$ (gauge fixing) we will obtain a "matter only" state of $B$ alone, with a global symmtery, corresponding to the dual model.

Since this seems very natural in the one dimensional setting, we will begin by introducing the procedure in $1+1d$. However, the procedure can be extended to higher space dimensions, where it can sometimes involve slightly unusual, but valid, gauging procedures (since there the dual sites are not associated with the links of the original lattice). We will show that too, by demonstrating with $2+1d$ cases.

\section{$\mathbb{Z}_N$ (clock) models in one space dimension}

Consider a one dimensional closed lattice (periodic boundary condition) with $M$ sites. Each lattice site $i\in\left\{0,...,M-1\right\}$ hosts an $N$ dimensional Hilbert space, on which we define two unitary operators, $Z_i$ and $X_i$, satisfying the $\mathbb{Z}_N$ algebra:
\begin{equation}\label{Z(N)algebra}
    \begin{matrix}
    Z_i^N=X_i^N=1, & Z_iZ_i^\dagger=X_iX_i^\dagger=1\\
    \\
    Z_iX_iZ_i^\dagger=e^{i\delta}X_i,& \delta=\frac{2\pi}{N}.
    \end{matrix}
\end{equation}
Operators associated with different sites commute. The spectrum of both types of operators contains all the $N$th roots of unity as non-degenerate eigenvalues; a conventional choice of basis is this of $Z$ eigenstates,
\begin{equation}
    Z\left|m\right\rangle = e^{i\delta m}\left|m\right\rangle.
\end{equation}

Using the $\mathbb{Z}_N$ algebra, one can show that the $X$ operators act as raising operators on $Z$ eigenstates, $X\left|m\right\rangle = \left|m+1\right\rangle$ (cyclically; $\left|N\right\rangle \equiv \left|0\right\rangle$). Similarly, the $Z$ opeartors act as lowering operators on $X$ eigenstates.

We can  represent the operators in the $Z$ eigenbasis as-
\begin{equation}\label{ZN:matrixform}
   \begin{matrix}
   Z=
    \left(\begin{matrix}
    1&0&&\\
    0&e^{i\delta}&&\\
    &&\ddots&\\
    &&&e^{i(N-1)\delta}
    \end{matrix}\right)
\\
\\
X=
 \left(\begin{matrix}
    0&0&0&...&1\\
    1&0&0&&&\\
    0&1&0&&&\\
  &&&\ddots  \\
    0&0&...&1&0
    \end{matrix}\right)
   \end{matrix}
   \end{equation}
- in the $\mathbb{Z}_2$ case this simply gives rise to $Z=Z^{\dagger}=\sigma_z$ and $Z=Z^{\dagger}=\sigma_x$.

The dynamics of clock models is given by a one parameter $(\lambda)$ family of Hamiltonians
\begin{equation}\label{HZN}
    H_A= \lambda \underset{i}{\sum}\left(Z_i+ Z^{\dagger}_i\right) +  \underset{i}{\sum}\left( X_i^\dagger X_{i+1}+X_{i+1}^\dagger X_i\right)
\end{equation}
this theory is also known as quantum clock model \cite{savit_duality_1980,horn_hamiltonian_1979}. For $N=2$, this is the quantum Ising model (in a trasversal field).

The Hamiltonian $H$ has a global $\mathbb{Z}_N$ symmetry,  manifested by the transformation  $X_i\to e^{i\delta}X_i$ and $Z_i\to Z_i$, $\forall i$. It is implemented by the unitary operator
\begin{equation}
    \mathcal{Z}=\prod_i Z_i.
\end{equation}
Note that $\mathcal{Z}^N=1$, and hence its eigenvalues are the $N$th roots of unity. 

One can verify that this is indeed a symmetry of the Hamiltonian: $\left[\mathcal{Z},H\right]=0$. As explained above, as a result of that, the Hilbert space factorizes into a direct sum of $N$ \emph{global} sectors, not mixed by the dynamics, which we labele by the eigenvalues of $\mathcal{Z}$. The matrix element of any $\mathbb{Z}_N$ symmetric operator between any two states from different sectors will vanish. The projection operator on the $q$th global selection sector (where $q\in \{0,1..,N-1\}$) is given by 
\begin{equation}\label{mprojection}
    {P}^{(q)}=\frac{1}{N}\sum_{n=0}^{N-1}\left(e^{-i\delta q} \mathcal{Z}\right)^n
\end{equation}

\subsection{Minimal Coupling of the Hamiltonian and Operators}
As anticipated, we would like to lift the global symmetry to a local one, that is
\begin{equation}
    X_i\to e^{i\delta m_i} X_i,
\end{equation}
where $m_i\in \{0,1..,N-1\}$,
through a conventional minimal coupling procedure.

While the local terms of the $\mathbb{Z}_N$ Hamiltonian are invariant under this transformation, the others, involving hopping of excitations between nearest neighbours, are not.
To fix that, we introduce new degrees of freedom, in the form of a $\mathbb{Z}_N$ gauge field, residing on the links (see Fig. \ref{fig:1d gauged}); on each link $i$ (labelled by its starting point) we introduce another $\mathbb{Z}_N$ Hilbert space. As the links are the sites of the dual lattice, this  introduces the gauge field Hilbert space, $\mathcal{H}_B$, which will be the one used for the dual system, eventually.

The new, minimally coupled, gauge invariant Hamiltonian reads 
\begin{equation}\label{H'ZN}
H'=\lambda\underset{i}{\sum}  \left(Z_i+Z^{\dagger}_i\right) + \underset{i}{\sum} \left( X_i^\dagger \tilde{Z}_i X_{i+1}+h.c. \right)
\end{equation}
where the gauge field operators, $\tilde{Z}_i$, defined on the  links connecting the sites $i$ and $i+1$, transform as
\begin{equation}
\begin{matrix}
    \tilde{Z}_i\to e^{i\delta m_i}\tilde{Z}_ie^{-i\delta m_{i+1}}
\end{matrix}
\end{equation}
under the local (gauge) transformations which are implemented by 
the unitaries
\begin{equation}\label{localZN}
    W_i=\tilde{X}_{i-1}Z_i\tilde{X}_i^\dagger.
\end{equation}

Normally, the physically relevant states of such a system $\mathcal{H}_{AB}= \mathcal{H}_A \times \mathcal{H}_B$ with a gauge invariant Hamiltonian, are the \emph{gauge invariant} ones - i.e., eigenstates of all symmetry transformations,
\begin{equation}
    W_i \left|\Psi\right\rangle_{AB} = e^{i\delta q_i} \left|\Psi\right\rangle_{AB}.
    \label{GaussLawZN}
\end{equation}
All these local constraints (one at each lattice site $i$), which we call \emph{Gauss laws}, split the Hilbert space of $\mathcal{H}_{AB}$ into further, local superselection sectors labelled by the eigenvalues $q_i\in\left\{0,...,N-1\right\}$  (static charges). Note that since 
\begin{equation}
\mathcal{Z}=\prod_i W_i,
\end{equation}
the global symmetry we had for $H$ alone still exists for the minimally coupled $H'$, and imposes some sector hierarchy on the  sectors: the sum $q =\underset{i}{\sum}q_i$ of the eigenvalues of all $W_i$ operators (modulo $N$), determines the global $\mathcal{Z}$ sector; or, phrased differenty, each global sector (determined by the eigenvalue of $\mathcal{Z}$ is a union of separate local sectors (determined by the eigenvalues of $W_i$).

\begin{figure}
    \centering
    \includegraphics[width=0.5\textwidth]{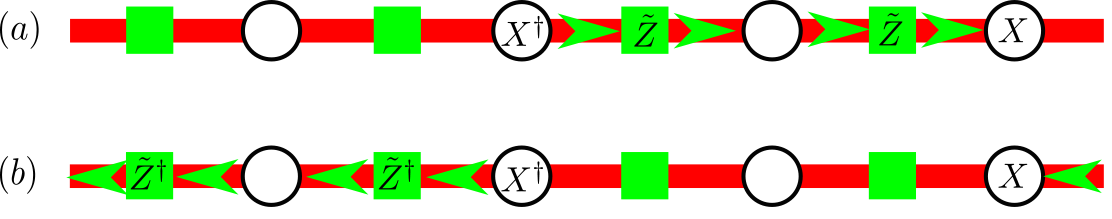}
    \caption{The two options to gauge in $1+1d$. }
    \label{fig:1d gauging}
\end{figure}

Just like we did for the Hamiltoian,  we can map any globally invariant operator $\mathcal{O}\in O\left(\mathcal{H}_A\right)$ to a locally, gauge invariant operator $\mathcal{O}'\in O\left(\mathcal{H}_A \times \mathcal{H}_B\right)$. 
The operators which respect the global symmetry and thus can be gauged are functions of the globally invariant operators $Z_i$, or products of the form $X^{n_{i_1}}_{i_1} X^{n_{i_2}}_{i_2} ...  X^{n_{i_K}}_{i_K}$ where $n_{i_1}+n_{i_2}+...+n_{i_K}=0$ (modulo $N$; $X^{-n}=X^{\dagger n}$), e.g.  pairs of the form $X_i^\dagger X_j$ for any $i$ and $j$. 
When gauging, we do not need to do anything with the $Z_i$ operators, since they are already gauge invariant. However, to gauge operators of the other kind, e.g. the pair $X_i^\dagger X_j$, we need to attach the sites $i,j$ by a flux string - a product of  $\tilde{Z}_i$ and/or  $\tilde{Z}^{\dagger}_i$ operators along a path connecting $i$ and $j$. In general, there are many choices for this path; in the Hamiltonian we picked the shortest path (and hence the name minimal coupling). In our periodic one dimensional setting, there are only two choices: suppose $i<j$; then we can gauge $X_i^\dagger X_j$ either to $X_i^\dagger \overset{j-1}{\underset{k=i}{\prod}}\tilde{Z}_k X_j$ or to $X_i^\dagger \overset{i-1}{\underset{k=j}{\prod}}\tilde{Z}^{\dagger}_k X_j$ (see Fig. \ref{fig:1d gauging}). We choose the first option, and will show that for the purpose of our procedure, we do not lose any generality by that.

\subsection{Embedding the $A$ States in the $AB$ Space}

Normally, minimal coupling is done on the level of Hamiltonian or Lagrangian, without considering the exact map of a globally invariant matter-only state to a locally invariant state of both gauge fields and matter.
This is since in most cases such an injective map cannot be defined: there are many possible gauge field configurations corresponding to the same matter configuration given by the original, globally invariant state; normally, the Gauss laws cannot be solved uniquely when considered as an equation for the gauge field \cite{emonts_gauss_2018}.

Since we would like to devise a map starting from unique states of the globally invariant system $A$, this might be a problem. However, it is not problematic at all thanks to the fact that our gauge invariant Hamiltonian $H'$ introduces no dynamics for the gauge field: all the $B$ system operators in $H'$ mutually commute - as these are only $\tilde{Z}_i$ and  $\tilde{Z}^{\dagger}_i$ opeartors. Normally, when performing a minimal coupling procedure one would add a pure-gauge term to $H'$, to introduce the missing dynamics, but this is not required in our case, since we only use the gauge symmetry  as a resource and we are not interested in enlarging the physical Hilbert space. We thus have two sets of operators mutually commuting among themselves and with the Hamiltonian,
\begin{equation}
    \left[W_i,W_j\right]=\left[W_i,H'\right]=0, \forall i,j,
\end{equation}
as well as
\begin{equation}
    \left[\tilde{Z}_i,\tilde{Z}_j\right]=\left[\tilde{Z}_i,H'\right]=0, \forall i,j;
\end{equation}
however, these sets of operators do not mutually commute with one another, and thus we can either consider the \emph{gauge invariant} states, eigenvalues of $W_i$, satisfying the Gauss laws (\ref{GaussLawZN}), as usual in gauge theories, or the \emph{gauge fixed} states, eigenvalues of $\tilde{Z}_i$, where the Gauss laws are violated and the gauge field has no dynamics. We choose the latter, since it allows us to easily embed a globally invariant state of $A$,  $\left|\psi\right\rangle_A \in \mathcal{H}_A$, in the extended $AB$ system, as a gauge fixed state. To see that, first consider the product of any eigenstate of $H$, $H\left|\phi\right\rangle = \epsilon \left|\phi\right\rangle$,
with the state 
\begin{equation}\label{in state}
    \left|\tilde{\text{in}}\right\rangle_B \equiv \underset{i}{\bigotimes}\left|\tilde 0\right\rangle
\end{equation}
(where $\tilde{Z}\left|\tilde{0}\right\rangle = \left|\tilde{0}\right\rangle$). The product state
$\left|\phi\right\rangle_A \otimes \left|\tilde{\text{in}}\right\rangle_B $ will not only be gauge fixed, it will also be an eigenstate of 
of $H'$ with the same energy, $\epsilon$. This is since
\begin{equation} \label{H'H}
    H'_{AB} \left|\tilde{\text{in}}\right\rangle_B = \left|\tilde{\text{in}}\right\rangle_B \otimes H_A.
\end{equation}
or
\begin{equation}
    H_A = \left\langle \tilde{\text{in}}\right| H'_{AB} \left|\tilde{\text{in}}\right\rangle
\end{equation}
- $\left|\tilde{\text{in}}\right\rangle_B$ is the gauge-fixing which completely "wipes off" the $B$ system from $H'$ to restore $H$.

Furthermore, for any gauged operator $\mathcal{O}'$, we have
\begin{equation}
    \mathcal{O}' \left|\tilde{\text{in}}\right\rangle_B = \left|\tilde{\text{in}}\right\rangle_B \otimes \mathcal{O}_A
    \label{O'O}
\end{equation} 
which proves why the choice of string for gauging is, as promised, not important in our case.

Thus, our first step in the duality transformation is to embed the globally invariant states $\left|\psi\right\rangle_A$ in the composite system,
\begin{equation}
\left|\psi\right\rangle_A \rightarrow \left|\psi\right\rangle_A \otimes \left|\tilde{\text{in}}\right\rangle_B  .  
\end{equation}
This embedding trivially preserves the observable physics,
\begin{equation}
    \left(\left\langle \psi' \right|_A \otimes \left\langle\tilde{\text{in}}\right|_B\right)
    \mathcal{O}' \left(\left|\psi \right\rangle_A \otimes \left|\tilde{\text{in}}\right\rangle_B\right) =
    \left\langle \psi' \right|\mathcal{O}\left|\psi \right\rangle
\end{equation}
for any globally invariant states $\left|\psi\right\rangle$,$\left|\psi'\right\rangle$.

\subsection{Switching the Roles of Matter and Gauge Fields}

Currently, we have two identical physical systems, $A$ on the sites and $B$ on the links (dual sites). They are both described by identical Hilbert spaces, but play two different physical roles: $A$ represents matter, and $B$ - gauge fields (we call this setting $AB$). But we could equally think of an opposite, or dual picture, in which $B$ represents matter, and $A$ represents the gauge fields (to which we will refer as $BA$). In the latter case, we expect gauge fixed states to be be eigenstates of the $Z_i$ operators, while gauge invariant states will be eigenstates of the \emph{dual gauge transformations},
\begin{equation}\label{dualZN}
    \tilde{W}_i=X_i\tilde{Z}_iX_{i+1}^\dagger
\end{equation}
(see Fig. \ref{fig:1d gauged}).

Can we map between these two settings? Is there a map which changes the roles of matter and gauge fields?
Ideally, we would like to find a unitary map $U$ between pairs of such operators. For example, we want it to map $W_i=\tilde{X}_{i-1}Z_i\tilde{X}_{i}$, the gauge constraint in the $AB$ setting (including $Z_i$ representing the matter), to $Z_i$ only, the gauge fixing constraint in the $BA$ setting (now representing the gauge field). 
We will construct the opposite map, $U^{\dagger}$, in two unitary steps: $Z_i\to \tilde{X}_i^\dagger Z_i\to  \tilde{X}_i^\dagger  \tilde{X}_{i-1}{Z}_i=W_i$, each involving two-body interactions only, of lattice sites with their closest dual sites.
For that, consider the $\tilde X_i$ eigenstates,
$\tilde{X}_i\left|\tilde{k}\right>_i=e^{i\tilde{k}\delta}\left|\tilde{k}\right>_i$, and note that 
\begin{equation}\label{Ustep1}
    Z_i\tilde{X}_i^\dagger=\sum_{\tilde{k}}Z_i e^{-i\delta \tilde{k}}\left|\tilde{k}\right>\left<\tilde{k}\right|_i
\end{equation}
Using (\ref{Z(N)algebra}) 
we can rewrite (\ref{Ustep1})  as
\begin{equation}
  Z_i\tilde{X}_i^\dagger=\left(\sum_{\tilde{k}}X_i^{\tilde{k}}\otimes\left|\tilde{k}\right>\left<\tilde{k}\right|_i\right)Z_i\left(\sum_{\tilde{k}'} X_i^{\dagger  \tilde{k}'}\otimes\left|\tilde{k}'\right>\left<\tilde{k}'\right|_i  \right)
\end{equation}

The second step is obtained similarly; we introduce
\begin{equation}
\begin{aligned}
    U_i=\sum_{\tilde{k}_i,\tilde{k }_{i-1}} \left(X_i^\dagger\right) ^{\tilde{k}_i-\tilde{k}_{i-1}}
    \otimes
    \left|\tilde{k}_{i-1}\right\rangle\left\langle\tilde{k}_{i-1}\right|_{i-1}
    \otimes
    \left|\tilde{k}_i\right\rangle
    \left\langle\tilde{k}_i\right|_i
\end{aligned}
    \end{equation}                                                                                                                  
for which
\begin{equation}
    W_i=\tilde{X}_i^\dagger  \tilde{X}_{i-1}{Z}_i=U_i^\dagger Z_iU_i
\end{equation}

Finally, we can express the switch operator as the product of all local building blocks $U_i$, 
\begin{equation}\label{unitaryopZn}
     U=\underset{i}{\prod} U_i
\end{equation} 

This transformation gives us what we wanted, and more:
\begin{equation}\label{Ureq}
\begin{matrix}
    U W_iU^\dagger=Z_i&,&U \tilde{Z}_i U^\dagger=\tilde{W}_i
\end{matrix}
.\end{equation}
That is, it also works in the other direction, mapping the gauge fixing conditions of the $AB$ setting to the Gauss laws of  $BA$.

$U$ switches the roles of the matter and the gauge fields, and thus we refer to it as the \emph{switch operator}. To demonstrate it, we can act with it on the state $\left|\psi\right\rangle_A \otimes \left|\tilde{\text{in}}\right\rangle_B$, which is gauge-fixed in the $AB$ setting, satisying $\tilde{Z}_i\left|\psi\right\rangle_A \otimes \left|\tilde{\text{in}}\right\rangle_B=\left|\psi\right\rangle_A \otimes \left|\tilde{\text{in}}\right\rangle_B$. After $U$ acts, we get a state satisfying the dual Gauss law at any site - eigenvalue of all the $\tilde{W}_i$ operators with no static charges (thanks to our choice of the state $\left|\tilde{\text{in}}\right\rangle$:
\begin{equation}\label{dualWsym}
\begin{aligned}
     \tilde{W}_iU \left|\psi \right\rangle_A \otimes \left|\tilde{\text{in}}\right\rangle_B &=U \tilde{Z}_i \left|\psi \right\rangle_A \otimes \left|\tilde{\text{in}}\right\rangle_B\\
     &= U\left|\psi \right\rangle_A \otimes \left|\tilde{\text{in}}\right\rangle_B.
\end{aligned}
\end{equation}
 
 On the other hand, if instead of the $AB$ gauge fixed state $\left|\psi\right\rangle_A \otimes \left|\tilde{\text{in}}\right\rangle_B$ we act with $U$ on an $AB$ gauge invariant state, the result will be a gauge fixed state in the $BA$ setting - an eigenstate of all $Z_i$ operators. This implies that if we can find a gauging transformation $W$, such that the state $W\left|\psi\right\rangle_A \otimes \left|\tilde{\text{in}}\right\rangle_B$ is gauge invariant in the $AB$ setting (satisfying the Gauss laws (\ref{GaussLawZN})), while containing the same physical information as $\left|\psi\right\rangle_A \otimes \left|\tilde{\text{in}}\right\rangle_B$ and therefore of $\left|\psi\right\rangle_A$, the state $UW\left|\psi\right\rangle_A \otimes \left|\tilde{\text{in}}\right\rangle_B$ will be of the form $\left|\text{out}\right\rangle_A\otimes\left|\tilde{\psi}\right\rangle_B$: a product of some trivial state in $A$ and a state in $B$ containing the same physics as $\left|\psi\right\rangle_A$ - the result of the desired duality transformation. Therefore we should construct the gauging operator $W$.

\subsection{The gauging operator}

To complete our argument, we introduce the gauging operator,
\begin{equation}\label{gaugingop}
    W\left(\left\{q_i\right\}\right) = \prod_{i}\frac{\sum_{n_i=0}^{N-1}\left(W_ie^{-i\delta q_i}\right)^{n_i}}{N}
\end{equation}
where $\{q_i\}$ is some static charge configuration $q_i\in\left\{0,...,N-1\right\}$ (similar gauging operators, without static charges, were used in \cite{haegeman_gauging_2015,zohar_eliminating_2018,zohar_removing_2019}). To gauge a state of the global sector $q$, we must make sure that $\underset{i}{\sum} q_i=q$ (modulo $N$). It is easy to note that
\begin{equation}
W_j W\left(\left\{q_i\right\}\right) = e^{i\delta q_j} W\left(\left\{q_i\right\}\right)
\end{equation}
and thus, the \emph{gauged state} we define by
\begin{equation}\label{gauged states}
   \left| \Psi\left(\left\{q_i\right\}\right)\right\rangle=\mathcal{N}^{-1/2}W\left(\left\{q_i\right\}\right)\left|\psi\right\rangle_A\otimes\left|\tilde{\text{in}}\right\rangle_B
\end{equation} 
is gauge invariant - satisfying the Gauss laws (\ref{GaussLawZN}) (the normalization factor $\mathcal{N}$ will be computed later):
\begin{equation}
    W_i \left| \Psi\left(\left\{q_i\right\}\right)\right\rangle = e^{i\delta q_i} \left|\Psi\left(\left\{q_i\right\}\right)\right\rangle
\end{equation}
 for every site $i$. 
 
 In fact, it is easy to see that $W\left(\left\{q_i\right\}\right)$ is a projection operator,  projecting the gauge fixed state $\left|\psi\right\rangle_A \otimes \left|\tilde{\text{in}}\right\rangle_B$ to a gauge invariant, local sector, with static charges $\left\{q_i\right\}$. 
This implies that it is non-unitary and in general not invertible, which might seem alarming since we want to use it in a process that preserves all the physical information. However, we will now show that it is an \emph{isometry}, and thus can be seen as unitary when restricted to the global sectors which is sufficient for us.

Let us begin our isometry proof by showing that inner products are preserved. Given a pair of states $\left|\psi_{\alpha}\right\rangle,\left|\psi_{\beta}\right\rangle$ in the same global symmetry sector of $A$, we would like to compute the overlap of the gauged states $\left|\Psi_{\alpha}\left(\left\{q_i\right\}\right)\right\rangle=\mathcal{N}^{-1/2}W\left(\left\{q_i\right\}\right)\left|\psi_{\alpha}\right\rangle_A\otimes\left|\tilde{\text{in}}\right\rangle_B$ and $\left|\Psi_{\beta}\left(\left\{q_i\right\}\right)\right\rangle=\mathcal{N}^{-1/2}W\left(\left\{q_i\right\}\right)\left|\psi_{\beta}\right\rangle_A\otimes\left|\tilde{\text{in}}\right\rangle_B$ (both in the same static charge configuration):
\begin{widetext}
\begin{equation}\label{normalization}
\begin{aligned}
\left\langle\Psi_\alpha \left(\left\{q_i\right\}\right)| \Psi_\beta  \left(\left\{q_i\right\}\right)\right\rangle &=  \frac{1}{\mathcal{N}} 
\left\langle\psi_\alpha\right|_A\otimes\left\langle\tilde{\text{in}}\right|_B W \left(\left\{q_i\right\}\right)
\left|\psi_\beta\right\rangle_A\otimes\left|\tilde{\text{in}}\right\rangle_B \\&=
         \frac{N^{-M}}{\mathcal{N}}\sum_{\{n_k\}}\left\langle\psi_\alpha\right|\prod_{j}\left(Z_je^{-i\delta q_j}\right)^{n_j} \left|\psi_\beta\right\rangle
         \left\langle\tilde{\text{in}}\right|\prod_{i}X_i^{n_{i+1}-n_i}\left|\tilde{\text{in}}\right\rangle
         \end{aligned},
\end{equation}
where M is the number of sites in lattice $A$, $n_i\in\{0,1..N-1\}$; and  $\left\langle\tilde{\text{in}}\right|\prod_{i}X_i^{n_{i+1}-n_i}\left|\tilde{\text{in}}\right\rangle$ will be nonzero if and only if, for all $i$, $n_i=n_{i+1}$. This implies that 
\begin{equation}
    \left\langle\Psi_{\alpha} | \Psi_{\beta} \right\rangle = \mathcal{N}^{-1} N^{-M} \left\langle\psi_{\alpha}\right|\underset{n}{\sum}\prod_{j}\left(Z_je^{-i\delta q_j}\right)^{n}\left|\psi_{\beta}\right\rangle=\mathcal{N}^{-1} N^{-M+1}\left\langle\psi_{\alpha} | \psi_{\beta} \right\rangle,
\end{equation} since        
\begin{equation}
\underset{n}{\sum}\underset{j}{\prod}\left(Z_je^{-i\delta q_j}\right)^{n}=P^{(\underset{j}{\sum}q_j)}
\end{equation}
is a projection operator onto the global symmetry sector into which these two states belong, as defined in (\ref{mprojection}). The inner product is thus preserved for the choice of normalization
        \begin{equation}
     \mathcal{N}=N^{-M+1}            \label{normaliation}
         \end{equation}
It is also straightforward to see that if the two states we begin with are gauged into different superselection sectors, the overlap will be zero since  $W\left(\left\{q_i\right\}\right)W\left(\left\{q'_i\right\}\right)=\underset{i}{\prod}\delta_{q_i,q'_i}W\left(\left\{q_i\right\}\right)$ (no summation over repeated indices), even if both local sectors belong to the same global one.  

Finally, let us consider the matrix elements of gauged operators, both to complete the isometry proof and to  make sure that physics is preserved by gauging: 
\begin{equation}\label{PO'P}
\begin{aligned}
    \left<\Psi_\alpha\left(\left\{q_i\right\}\right)\right| \mathcal{O}' \left|\Psi_\beta\left(\left\{q_i\right\}\right)\right>&=
    N^{M-1}
    \left\langle\psi_\alpha\right|_A\otimes\left\langle\tilde{\text{in}}\right|_B W\left(\left\{q_i\right\}\right) \mathcal{O}' W\left(\left\{q_i\right\}\right)
\left|\psi_\beta\right\rangle_A\otimes\left|\tilde{\text{in}}\right\rangle_B 
\\&=N^{M-1}
    \left\langle\psi_\alpha\right|_A\otimes\left\langle\tilde{\text{in}}\right|_B W\left(\left\{q_i\right\}\right) \mathcal{O}' 
\left|\psi_\beta\right\rangle_A\otimes\left|\tilde{\text{in}}\right\rangle_B 
\\&=N^{M-1}
    \left\langle\psi_\alpha\right|_A\otimes\left\langle\tilde{\text{in}}\right|_B W\left(\left\{q_i\right\}\right) \left(\mathcal{O}
\left|\psi_\beta\right\rangle_A\right)\otimes\left|\tilde{\text{in}}\right\rangle_B 
= \left\langle\psi_\alpha\right| \mathcal{O} \left|\psi_\beta\right\rangle
\end{aligned}
\end{equation}
\end{widetext}
Using the fact that $\mathcal{O}'$, as a gauge invariant operator, commutes with $W$, as a linear combination of gauge transformations and  Eq. (\ref{O'O}) which allowed us to convert $\mathcal{O}'$ to $\mathcal{O}$ when acting on $\left|\tilde{\text{in}}\right\rangle$. Finally, using similar arguments to those used for the inner product, one can see that $N^{M-1}\left\langle \tilde{\text{in}}\right| W\left(\left\{q_i\right\}\right) \left|\tilde{\text{in}}\right\rangle = P^{(\underset{i}{\sum}q_i)}$: the projector onto the relevant global symmetry sector, and thus indeed the matrix elements are preserved, the transformation is an isometry and the physics is not harmed by gauging.

\subsection{The dual picture}
Starting from the globally invariant operators $\mathcal{O}$ acting on $\mathcal{H}_A$, we defined the primed operators $\mathcal{O'}$ acting on $\mathcal{H}_A \times \mathcal{H}_B$ which are gauge invariant in the $AB$ sense. Using the switch operator $U$ from (\ref{unitaryopZn}), we can define their dual partners,
\begin{equation}\label{odual}
    \tilde{\mathcal{O}}'=U\mathcal{O}'U^\dagger
.\end{equation}
For example, 
\begin{equation}\label{dualprimeH}
    \tilde{H}'_{BA}=UH'_{AB}U^\dagger=\lambda\sum_i  \left(\tilde{X}_{i-1}^\dagger Z_i\tilde{X}_i+h.c .\right) +\sum_i\left(\tilde{Z}_i+\tilde{Z}_i^\dagger\right)
.\end{equation}

As expected, the operators $\tilde{\mathcal{O'}}$  defined in (\ref{odual}) are gauge invariant in the $BA$ setting (commute with the local symmetry transformations in the $BA$ setting, $\tilde{W}_i$, which can be shown using $\left[\tilde{Z}_i,\mathcal{O}'\right]=0$ as well as Eqs. (\ref{odual}) and  (\ref{Ureq}). 
Similarly, one can show that $\left[Z_i,\tilde{\mathcal{O}}'\right]=0$ for all $i$, and in addition, 
$\left[Z_i,\tilde{H}'\right]=0$. Therefore, as discussed above, the gauge field fsystem, this time system $A$, has no dynamics.

The action of the switch operator on the gauged state can be written as
\begin{widetext}
\begin{equation}\label{UPSIW}
\begin{aligned}
U\left|\Psi\left(\{q_i\}\right)\right\rangle&=\mathcal{N}^{-\frac{1}{2}}UW\left(\{q_i\}\right)U^\dagger U\left|\psi\right\rangle_A\otimes\left|\tilde{\text{in}}\right\rangle_B=
    \mathcal{N}^{-\frac{1}{2}}\underset{i}{\prod}\Pi_{q_i}(i)U\left|\psi\right\rangle_A\otimes\left|\tilde{\text{in}}\right\rangle_B\equiv\left(\underset{i}{\bigotimes}\left|q_i\right\rangle_A\right)\otimes\left|\tilde{\psi}\right\rangle_B
\end{aligned}
\end{equation}
\end{widetext}
using Eq. (\ref{Ureq}) and
\begin{equation}\label{UWU}
UW\left(\{q_i\}\right)U^\dagger = \underset{i}{\prod}\frac{\sum_{n_j=0}^{N-1} \left(Z_ie^{-iq_i}\right)^{n_j}}{N}=\underset{i}{\prod}\left|q_i\right\rangle\left\langle q_i\right|_i,
\end{equation} 
  where ${Z}_i\left|q_i\right\rangle_i = e^{i\delta q_i}\left|q_i\right\rangle_i$. The resulting state is a product between the trivial gauge fixed state $\underset{i}{\bigotimes}\left|q_i\right\rangle_A$ of $A$ and some state of $B$, disentangled from it, which we denote by $\left|\tilde{\psi}\right\rangle_B$.

Choosing the configuration with all $q_i=0$ (no static charges), we define the "out" state 
\begin{equation}\label{out}
    \left|{\text{out}}\right\rangle_A \equiv \underset{i}{\bigotimes}\left| 0\right\rangle
\end{equation}
- that is, a product state of the ${Z}_i$ eigenstates with eigenvalue $1$ on all the sites.
Since $\left[Z_i,\tilde{\mathcal{O}}'\right]=0$ for all $i$ as argued above, for any $\tilde{\mathcal{O'}}$ there exists an operator $\tilde{\mathcal{O}}_B$ acting on $\mathcal{H}_B$ alone, such that
\begin{equation}\label{dualeigenoperator}
    \tilde{\mathcal{O}}' \left|{\text{out}}\right\rangle_A = \left|{\text{out}}\right\rangle_A \otimes \tilde{\mathcal{O}}_B
\end{equation}

With all the building blocks at hand, it is only left for us now to conclude and show  how to relate the physics of $\left|\psi\right\rangle_A$ and $\left|\tilde{\psi}\right\rangle_B$. 
In (\ref{PO'P}) we have shown that 
$\left<\Psi_\alpha\right|\mathcal{O}' \left|\Psi_\beta\right>=\left\langle\psi_\alpha\right| \mathcal{O} \left|\psi_\beta\right\rangle$. Using (\ref{odual}), (\ref{UPSIW}) and (\ref{dualeigenoperator}) we can proceed and obtain   
\begin{widetext}
\begin{equation}\label{TPO'TP}
    \left\langle\psi_\alpha\right| \mathcal{O} \left|\psi_\beta\right\rangle=\left<\Psi_\alpha\right|U^\dagger \tilde{\mathcal{O}}' U \left|\Psi_\beta\right>=\left\langle\text{out}\right|_A\otimes\left\langle\tilde{\psi}_\alpha\right|_B \tilde{\mathcal{O}}'  \left|\text{out}\right\rangle_A\otimes\left|\tilde{\psi}_\beta\right\rangle_B\\
    \\=\left<\tilde{\psi}_\alpha\right| \tilde{\mathcal{O}} \left|\tilde{\psi}_\beta\right>
\end{equation}

Indeed, system $B$, at the end of the process, describes the dual theory. Its Hamiltonian is obtained from (\ref{dualprimeH}), following the steps above, as
\begin{equation}\label{dualH}
    \tilde{H}_B=\left\langle\text{out}\right|\tilde{H}'_{BA}\left|\text{out}\right\rangle =\lambda\sum_i  \left(\tilde{X}_{i-1}\tilde{X}_i^\dagger+h.c .\right) +\sum_i\left(\tilde{Z}_i+\tilde{Z}_i^\dagger\right)
.\end{equation}
\end{widetext}
This Hamiltonian is related to the Hamiltonian of system $A$ by $ H\left(\mathcal{O},\lambda\right)=\lambda\tilde{H}\left(\tilde{\mathcal{O}},\lambda^{-1}\right)$, manifesting the well known self-duality of the Ising and clock models in a single space dimension \cite{fradkin_order_1978,savit_duality_1980}, which shows that the strong coupling behaviour $(\lambda<1)$ and weak coupling behaviour $(\lambda>1)$ are in a sense equivalent. ay be inverted.

\subsection{Feasibility}

Let us summarize what we have so far achieved. We started with an arbitrary state at system $A$, $\left|\psi\right>_A$, respecting the global symmetry. We embedded it in the extended system $AB$, introducing the gauge-fixed state $\left|\psi\right\rangle_A \otimes \left|\tilde{\text{in}}\right\rangle_B$; this state preserved the physics in terms of the gauged observables $\mathcal{O'}$, but was gauge fixed. We further gauged the state using the gauging operator, $W$, which, as we showed, is an isometry. Finally, we used the switch operator $U$ which changed the roles of the gauge fields and the matter fields, taking us from a gauge invariant state in the $AB$ sense to a gauge fixed state $\left|\text{out}\right\rangle_A \otimes \left|\tilde{\psi}\right\rangle_B$ in the $BA$ sense, where $\left|\text{out}\right\rangle_A \otimes \left|\tilde{\psi}\right\rangle_B$ contains all the physical information of $\left|\psi\right>_A$, but in a dual formulation. All that can be summarized by
\begin{widetext}
\begin{equation}
    \left|\psi\right\rangle_A \rightarrow \left|\psi\right\rangle_A \otimes \left|\tilde{\text{in}}\right\rangle_B  \rightarrow 
       \mathcal{N}^{-1/2}W\left|\psi\right\rangle_A \otimes \left|\tilde{\text{in}}\right\rangle_B
         \rightarrow U\mathcal{N}^{-1/2}W\left|\psi\right\rangle_A \otimes \left|\tilde{\text{in}}\right\rangle_B  
    \rightarrow \left|\text{out}\right\rangle_A \otimes \left|\tilde{\psi}\right\rangle_B
    \rightarrow  \left|\tilde{\psi}\right\rangle_B
\end{equation}
\end{widetext}
This is a physical rather than mathematical description of the concept of a duality transformation. Instead of linking the two formulations dual to one another by performing some mathematical operations, we actually introduced a second physical system, and using physical operations transferred all the physical  information to it. This can be usfeul for quantum simulation, where one may wish to alternate between the two dual formulations of the same model, for example in order to better simulate experimentally different coupling regimes (later on, when we discuss two space dimensions, we will return to this issue). For that, we only need to have the $B$ system ready and properly coupled, and then implement the $W$ and $U$ operations. 

While the switch operator $U$ is a simple local operation which can be constructed out of two-body interaction terms and hence in principle feasible on quantum devices (recall the steps we introduced for its construction), the $W$ operation is more challenging to implement. However, note that one does not actually have to use it explicitly: thanks to (\ref{UWU}), we see that instead of gauging and then acting with the switch operator, one can first act with the switch operator $U$ and afterwards, using simple local measurements, post-select the state of the original system to $\left|\text{out}\right\rangle$. This way, the exact same  state $\left|\tilde{\psi}\right\rangle$ is obtained in the system $B$. Thus, one can transfer the physics to the dual picture by properly introducing $\left|\tilde{\text{in}}\right\rangle$, acting with the local unitary $U$ and post selecting the state of $A$ to $\left|\text{out}\right\rangle$ - using  feasible operations only, involving two-body interactions at most. The procedure can then be summarized by
\begin{widetext}
\begin{equation}
    \left|\psi\right\rangle_A \rightarrow \left|\psi\right\rangle_A \otimes \left|\tilde{\text{in}}\right\rangle_B  \rightarrow 
       U\left|\psi\right\rangle_A \otimes \left|\tilde{\text{in}}\right\rangle_B
         \rightarrow  \left|\text{out}\right\rangle_A \left\langle\text{out}\right|_A U\left|\psi\right\rangle_A \otimes \left|\tilde{\text{in}}\right\rangle_B 
    \rightarrow \left|\text{out}\right\rangle_A \otimes \left|\tilde{\psi}\right\rangle_B
      \rightarrow  \left|\tilde{\psi}\right\rangle_B
\end{equation}
\end{widetext}

Similar procedures with such $U$ operators have been previously used in proposals for quantum simulation of lattice gauge theories, using the stator formalism \cite{reznik_remote_2002,zohar_half_2017} which allows one to map many-body interactions to sequences of two-body interactions of each physical degree of freedom with an auxiliary one. This operation is nothing but the switch operator, when the ancillary system is the dual one. In this context, methods to implement the switch operators were introduced for cold atomic systems \cite{zohar_digital_2017-1,zohar_digital_2017-1,bender_digital_2018} (only $U$, without the last step of post-selection to the dual state) as well as very recently nanophotonic or cavity-QED setups \cite{armon_photon_2021}, where post-selection to the dual state was used. 

Note that from a physical point of view, everything we have done so far may be extended to higher dimensions too, as we shall do in the following parts of this work; however, the physical concept of the remaining parts are exactly the same.

\subsection{Invertibility}

The procedure introduced above was proven to be an isometry between global sectors of $\mathcal{H}_A$ and $\mathcal{H}_B$, and thus it can be simply inverted, using the inverse switch operator $U^{\dagger}$ and a gauging operator $\tilde{W}\left(\tilde{q}_i\right)$ for the $BA$ setting, defined similarly to $W\left(q_i\right)$ from Eq. (\ref{gaugingop}). This can be summarized by
\begin{widetext}
\begin{equation}
    \left|\tilde{\psi}\right\rangle_B \rightarrow 
   \left|\text{out}\right\rangle_A \otimes \left|\tilde{\psi}\right\rangle_B \rightarrow 
   \mathcal{N}^{-1/2}\tilde{W}\left|\text{out}\right\rangle_A \otimes \left|\tilde{\psi}\right\rangle_B
   \rightarrow 
   U^{\dagger}\mathcal{N}^{-1/2}\tilde{W}\left|\text{out}\right\rangle_A \otimes \left|\tilde{\psi}\right\rangle_B
    \rightarrow
    \left|\psi\right\rangle_A \otimes \left|\tilde{\text{in}}\right\rangle_B  \rightarrow  
   \left|\psi\right\rangle_A
\end{equation}
or
\begin{equation}
    \left|\tilde{\psi}\right\rangle_B \rightarrow 
   \left|\text{out}\right\rangle_A \otimes \left|\tilde{\psi}\right\rangle_B \rightarrow 
   U^{\dagger}\left|\text{out}\right\rangle_A \otimes \left|\tilde{\psi}\right\rangle_B
    \rightarrow 
    \left|\tilde{\text{in}}\right\rangle  \left\langle\tilde{\text{in}}\right|_BU^{\dagger}\left|\text{out}\right\rangle_A \otimes \left|\tilde{\psi}\right\rangle_B
    \rightarrow
    \left|\psi\right\rangle_A \otimes \left|\tilde{\text{in}}\right\rangle_B  \rightarrow  
   \left|\psi\right\rangle_A
\end{equation}
\end{widetext}

In the one dimensional case discussed so far, the model was self-dual, and thus the gauging transformations have the same functional form. In other cases - for example in the $2+1d$ clock model which we study below - since there is no self duality. The gauging procedures of both directions are thus not the same, but can be defined.

\section{ $2+1$d $\mathbb{Z}_N$ models}\label{2DZN}

Consider a two dimensional square spatial lattice on a torus (periodic boundary conditions again), with unit vectors denoted by
  $\{\hat{1},\hat{2}\}$, where $\hat{1}$ points to the right and $\hat{2}$ upwards, respectively. We will walk through the implementation of the duality between clock models - the straightforward generalization of the one dimensional case, and their dual models, pure $\mathbb{Z}_N$ lattice gauge theories: the models are not self dual, making the process more interesting \cite{fradkin_order_1978,fradkin_phase_1979,horn_hamiltonian_1979,elitzur_phase_1979}.
  
  In this section, we will focus on generalizing our gauging procedure, introduced above, for this two dimensional case. Conceptually, the process is very similar, and sometimes identical; we will thus focus only on the aspects which differ from the one dimensional case.
  
  \subsection{From the Clock Model to a $\mathbb{Z}_N$ Lattice Gauge Theory}
  
  \subsubsection{The Model}
  We consider, once again, the $\mathbb{Z}_N$ clock model for the physical system $A$. On each lattice site, denoted by $i$, we place the familiar $\mathbb{Z}_N$ Hilbert space, and consider the Hamiltonian 
\begin{equation}\label{2dclockmodel}
H=\underset{i}{\sum}\lambda\left( Z_i+Z_i^\dagger\right)+\underset{i;k=1,2}{\sum} \left(X_{i}^\dagger X_{i+\hat{k}}+h.c.\right)
\end{equation}

We would like to find the dual picture of this theory, using the procedure established above.
The global $\mathbb{Z}_N$ symmetry of the system $A$ is a straightforward generalization of what we had in the one dimensional case, and it is implemented by the unitary operator
\begin{equation}\label{2dglobalZ}
    \mathcal{Z}=\underset{i}{\prod}Z_i.
\end{equation}
$\mathcal{Z}$ commutes with the two dimensional Hamiltonian $H$ (\ref{2dclockmodel}), hence splitting $\mathcal{H}_A$ to $N$ global selection sectors, labelled by $q=\left\{0,...,N-1\right\}$.

\subsubsection{Minimal Coupling}
In the $AB$ setting, the $A$ degrees of freedom represent the matter, and we look for a way to lift the global symmetry to a local one, acting on the matter as 
 $X_i\to e^{i\delta m_i}X_i$ and $Z_i\to Z_i$. As usual, the Hamiltonian (\ref{2dclockmodel}) is not invariant under this transformation, due to the nearest-neigbour term, and to make it invariant we need to introduce $\mathbb{Z}_N$ gauge fields, residing on the links - the system $B$. This picture is the $AB$ setting: where $A$ represents matter and $B$ gauge fields. 
The minimally coupled Hamiltonian is
\begin{widetext}
\begin{equation}\label{2d gauged clock-model}
H'=\lambda\underset{i}{\sum}\left( Z_i+Z_i^\dagger\right)+\underset{i}{\sum} \left(X_{i}^\dagger \tilde{Z}_{i;1}^{\dagger} X_{i+\hat{1}}+X_{i}^\dagger \tilde{Z}_{i;2} X_{i+\hat{2}}+h.c.\right)
\end{equation}
\end{widetext}
where 
$\tilde{Z}_{i;k}$ is a gauge field operator, acting on the link emanating from $i$ in direction $k$ (connecting the sites $i$ and $i+\hat{k}$.
\begin{figure}
    \centering
    \includegraphics[width=0.3\textwidth]{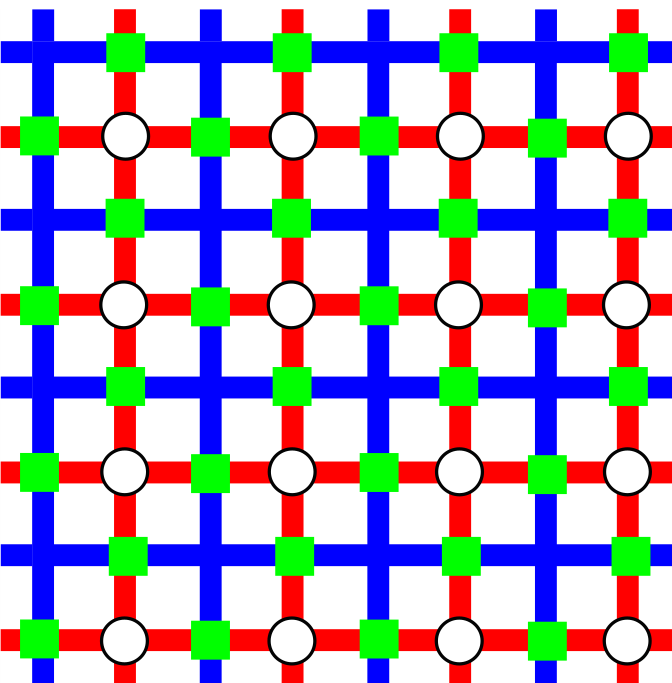}
    \caption{The two dimensional lattice. Lattice $A$, of the original theory, is shown in red, while the dual lattice, $B$, is blue. Sites of the $A$ lattice, where the original degrees of freedom reside, are denoted by black circles, and they coincide with the plaquettes of the dual lattice $B$. The $B$ degrees of freedom reside on the links of both lattices, and are denoted by green squares. }
    \label{fig:gauged 2d system}
\end{figure}

This choice of gauging might seem unusual (with  $\tilde{Z}$ and $\tilde{Z}^{\dagger}$ imbalance in both directions). We made this choice  to obtain something simpler and recognizable at the end of the process, when everything is denoted using terms of the $B$ lattice, which is the dual lattice of 
 of $A$ (see Fig. \ref{fig:gauged 2d system}), which we shall do already now.

\begin{figure*}
    \centering
    \includegraphics[width=0.7\textwidth]{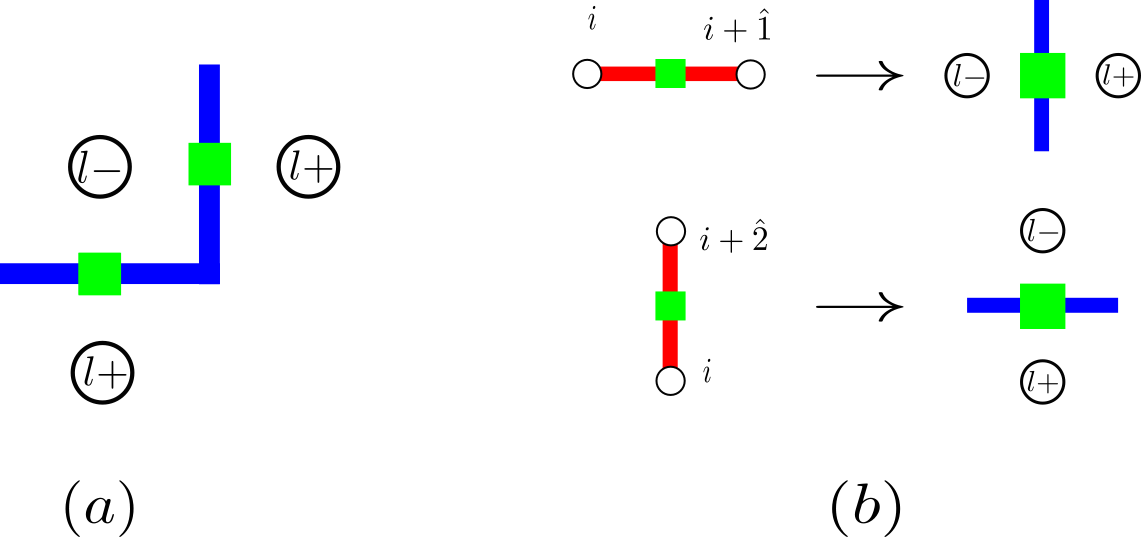}
   \caption{(a) Notation convention for $l\pm$. Black circles represent lattice sites of system $A$, or plaquettes of the dual system $B$. (b) Change of notation from the original lattice ($A$)) to its dual ($B$). The $A$ links are red, while the $B$ ones are blue. he $A$ degrees of freedom reside on the $A$ sites, $i$, which are the plaquettes of $B$, denoted by $l_{\pm}$ as defined in (a); a horizontal $A$ link becomes a vertical $B$ link and vice-versa. The $B$ degrees of freedom (green squares) are on the links of both lattices. }
   \label{fig:notations}
\end{figure*}

The links of $A$ coincide with those of $B$, but in orthogonal directions, while the sites of $A$ coincide with the plaquettes of the $B$ lattice. Denoting the plaquette of $B$ by $p$, and the links of $B$ by $l$, we can rewrite the minimally coupled Hamiltonian as
\begin{equation}\label{2dprimeclockmodel}
\begin{aligned}
H'_{AB}=\lambda\underset{p}{\sum}\left(
Z_{p}+Z^\dagger_p\right)+\underset{l}{\sum} \left( X_{l+}^\dagger \tilde{Z}_{l}X_{l- }
+h.c.\right)
\end{aligned}
,\end{equation}
where $l_-$ refers to the $B$ plaquette below or on the right of the link $l$, and $l_+$ to the plaquette on its right or below it (see Fig. \ref{fig:notations}).

\begin{figure}[h]
\includegraphics[width=0.15\textwidth]{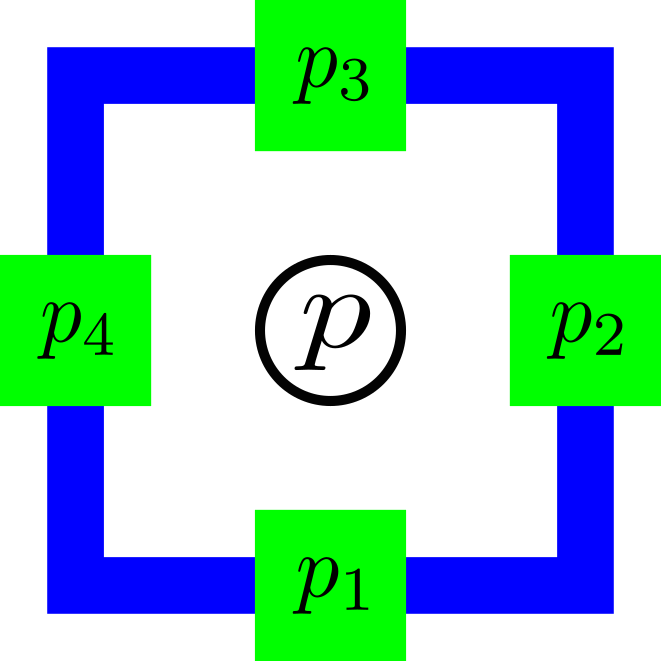}
    \caption{Notation convention for the links of the dual lattice around the site $p$.}
    \label{fig:plaquette}
\end{figure}

\begin{figure*}
    \centering
    \includegraphics[width=\textwidth]{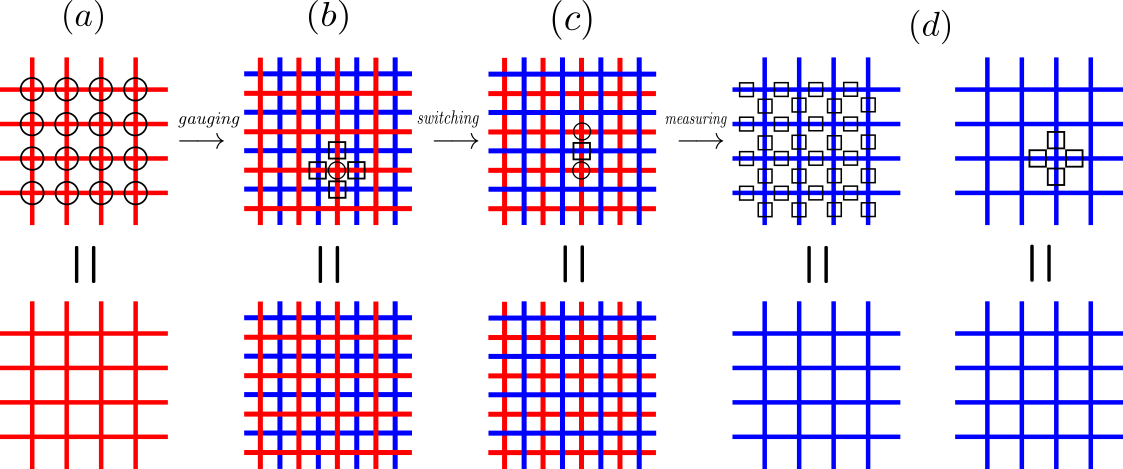}
    \caption{symmetries of each of the theories: (a) system $A$ is described by a red lattice; its degrees of freedom  reside on the sites. It has a global symmetry: acting with the same symmetry operation on all the sites (top) leaves the state invariant (bottom).  (b) the composite system in the $AB$ sense has, on top of the global symmetry of $A$, a local symmetry involving a single $A$ matter component and four $B$ gauge field links around it. (c) switching to the $BA$ settings, we have another gauge symmetry, involving two $A$ components (gauge fields) and one $B$ (matter). (d) in the $B$ setting we have both a global and a local, pure gauge transformation.  }
    \label{fig:2d symmetries}
\end{figure*}

The gauge field operator $\tilde{Z}_l$ transforms as:
\begin{equation}
   \tilde{Z}_l\to e^{i\delta m_{l+}} \tilde{Z}_l e^{-i\delta m_{l-}} 
;\end{equation}
in terms of the $A$ ($B$) lattice, the gauge transformation acts on a site (plauqette) and all the links around it  (see Fig.\ref{fig:2d symmetries}). To express things in a simple way using the dual lattice, we denote the dual links around the dual plaquette $p$ by $p_1,p_2,p_3$, and $p_4$, in a conuter-clockwise order from the lower link on (see 
Fig.\ref{fig:plaquette}).

Then it is very simple to write down the local unitaries  
  \begin{equation} \label{Wpdef}
    W_{p}=Z_{p}\tilde{X}_{p_1}\tilde{X}_{p_2}\tilde{X}_{p_3}^\dagger\tilde{X}_{p_4}^\dagger
\end{equation}
which implement the gauge transformations; indeed $\left[W_p,H'_{AB}\right]=0$ for every plaquette $p$ of the dual lattice $B$.

Just as in the one dimensional case, the gauged Hamiltonian $H'$ contains no dynamics for the gauge fields $B$, since $\left[H',\tilde{Z}_l\right]=0$ for any $l$. We define the product state
\begin{equation}
 \left|\tilde{\text{in}}\right\rangle_B = \underset{l}{\bigotimes}\left|\tilde{0}\right\rangle_l.   
    \end{equation}
For any eigenstate of $H$, $\left|\psi\right\rangle$,  $\left|\psi\right\rangle_A \otimes \left|\text{in}\right\rangle_B$ 
will be an eigenstate of $H'$ with the same energy, since the relation (\ref{H'H}) is satisfied in two dimensions too. Thus the first step is, again, the embedding
\begin{equation}
\left|\psi\right\rangle_A \rightarrow \left|\psi\right\rangle_A \otimes \left|\text{in}\right\rangle_B 
\end{equation}

\begin{figure}[h]
    \centering
    \includegraphics[width=0.3\textwidth]{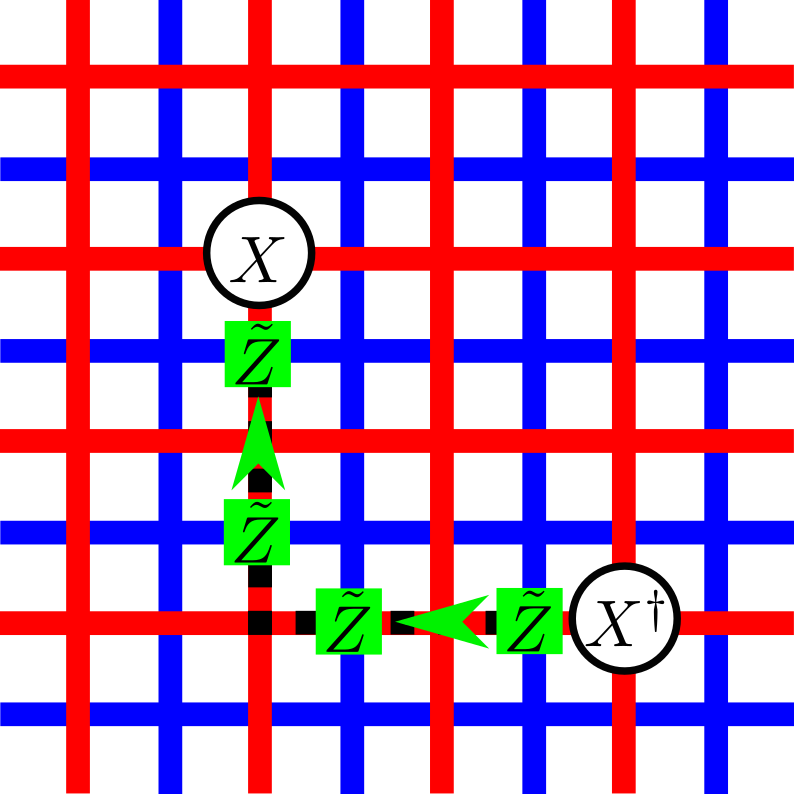}
    \caption{Gauging $X_p^{\dagger} X_p'$ involves attaching a string of $\tilde{Z}$ or $\tilde{Z}^{\dagger}$ operators along any path $\mathcal{C}$ connecting them. The choice between $\tilde{Z}$ and $\tilde{Z}^{\dagger}$ is made by the link's orientation (parallel or anti-parallel to to the unit vectors, respectively): the string emanates from positive charges ($X^{\dagger}_p$) and enters negative ones ($X_p'$). One can make several choices for the path $\mathcal{C}$, but this is not important in our case, as explained in the text.  }
    \label{fig:gauging}
\end{figure}

One also needs to specify a prescription for gauging any other globally invariant operator $\mathcal{O}_A$ to $\mathcal{O}'_{AB}$. This is a simple generalization of the one dimensional case too, by attaching strings of $\tilde{Z}$ and $\tilde{Z}^{\dagger}$ operators where necessary. Consider, for example, operators of the form $X^\dagger_p X_{p'}$ for some pair of $B$ plaquettes ($A$ sites). In the one dimensional case we had two ways of connecting them, and here we have much more -  any operator of the form $X_p^\dagger \underset{l\in \mathcal{C}}{\prod}\tilde{Z}_l X_{p'}$ where $\mathcal{C}$ is any path along the lattice connecting $p$ and $p'$ ($\tilde{Z}_l$ may have to be replaced by its hermitian conjugate on some links, depending on the loop's orientation) - see Fig. \ref{fig:gauging}. However, just like in the one dimensional case, since we start with the gauge fixed states $\left|\psi\right\rangle_A \otimes \left|\text{in}\right\rangle_B$ we can make any choice we wish.

\subsubsection{The Switch Operator}
Next, we construct the $2+1$d switch operator, which switches roles between the gauge and matter degrees of freedom, taking us from the $AB$ to the $BA$ setting, and from gauge invariance to gauge fixing. Even if we had no clue what the dual model was, we could still find it: it has to be some unitary $U$, mapping the Gauss law operators of the $AB$ setting, $W_p$ from (\ref{Wpdef}) to some gauge fixing constraint in the $BA$ setting, thus on the $A$ system. Looking at the form of $W_p$, we see that the only $A$ contribution to it is $Z_p$, and thus we would like to find a transformation mapping $W_p$ to $Z_p$.

A straightforward generalization of the one dimensional procedure allows us to define (again as a product of two-body operations)
  \begin{widetext}
  \begin{equation}\label{U2d}
    U_{p}=\underset{
    \tilde{k}_{p_1},\tilde{k}_{p_2}\tilde{k}_{p_3},\tilde{k}_{p_4}
    }{\sum}  \left(X_{p}\right)^{\tilde{k}_{p_1}+\tilde{k}_{p_2}-\tilde{k}_{p_3}-\tilde{k}_{p_4}} \otimes
    \left|\tilde{k}_{p_1}\right\rangle\left\langle\tilde{k}_{p_1}\right|_{p_1} \otimes
    \left|\tilde{k}_{p_2}\right\rangle\left\langle\tilde{k}_{p_2}\right|_{p_2} \otimes
    \left|\tilde{k}_{p_3}\right\rangle\left\langle\tilde{k}_{p_3}\right|_{p_3} \otimes
    \left|\tilde{k}_{p_4}\right\rangle\left\langle\tilde{k}_{p_4}\right|_{p_4}
,\end{equation}
  \end{widetext}
   where $\tilde{X}_{p_i}\left|\tilde{k}_{p_i}\right\rangle_{p_i}=e^{i\delta k_{p_i}}\left|\tilde{k}_{p_i}\right\rangle_{p_i}$, and $k_p\in\{0,...,N-1\}$ (as before, this can be easily obtained from a sequence of two body interactions). We define
    $U=\underset{p}{\prod}U_{p}$, and, as required, it gives rise to  
    \begin{equation}
U W_p U^{\dagger} = Z_p.
    \end{equation}
for every $p$. Besides that, one can show that
\begin{equation}
U \tilde{Z}_l U^{\dagger} = X^{\dagger}_{l-} \tilde{Z}_l X_{l+}  
\end{equation}
and since $\tilde{Z}_l$ is the gauge fixing operator of the $AB$ setting, we deduce that the Gauss law operators of the $BA$ setting are 
\begin{equation}\label{tildeWl}
    \tilde{W}_l = X^{\dagger}_{l-} \tilde{Z}_l X_{l+}  
\end{equation}
- functionally different from $W_p$, giving us a hint that we are not dealing with a self-dual model in this case (even if we did not know it already). In the $AB$ setting, the gauge transformations act on a site of $A$ (or a plaquette of $B$) and the links around it (gauge fields). - altogether on five Hilbert spaces, while in the $BA$ settings, they act on a link of $B$ (matter) and the two plaquettes on its sides ($A$ sites, gauge fields).

By applying $U$ onto the $AB$ Hamiltonian $H'$ (\ref{2dprimeclockmodel}), we obtain the switched, $BA$ Hamiltonian
\begin{equation}\label{tildeHBA}
\begin{aligned}
    \tilde{H}'_{BA}&=UH'_{AB}U^\dagger\\&=
   \underset{l}{\sum}\left(\tilde{Z}_l+\tilde{Z}_l^\dagger\right)+
    \lambda\underset{p}{\sum}\left( \tilde{X}_{p_1}^\dagger \tilde{X}^\dagger_{p_2} \tilde{X}_{p_3} \tilde{X}_{p_4}Z_p +h.c.\right)
    \end{aligned}.
\end{equation}
As expected, it commutes with all the $\tilde{W}_l$ and $Z_p$ operators (allowing to fix the gauge), giving rise to \begin{equation}\label{2ddualH}
\begin{aligned}
    \tilde{H}_B&=\left\langle\text{out}\right|\tilde{H}'_{BA}\left|\text{out}\right\rangle\\&=
   \underset{l}{\sum}\left(\tilde{Z}_l+\tilde{Z}_l^\dagger\right)+
    \lambda\underset{p}{\sum}\left( \tilde{X}_{p_1}^\dagger \tilde{X}^\dagger_{p_2} \tilde{X}_{p_3} \tilde{X}_{p_4} +h.c.\right).
\end{aligned}
    \end{equation}
Where $\left|\text{out}\right\rangle = \underset{p}{\bigotimes}\left|0\right\rangle_p$.

\subsubsection{The gauging operator}

Suppose the state $\left|\psi\right\rangle_A$ belongs to the global sector $q$. Then, the embedded, gauge-fixed state  $\left|\psi\right\rangle_A \otimes \left|\tilde{in}\right\rangle_B$ is gauged using
\begin{equation}
    W\left(\{q_{p}\}\right)=\prod_{p}\frac{\sum_{n=0}^{N-1}\left(W_{p}e^{-i\delta q_{p}}\right)^n}{N},
\end{equation}
for any static charge configuration $\{q_{p}\}$  belonging to the right global sector, that is $\sum_pq_p=q$ (modulo $N$).
The normalized gauged state is given by 
\begin{equation} \label{2dgauged}
    \left|\Psi\left(\left\{q_p\right\}\right)\right\rangle=\mathcal{N}^{-1/2}W\left(\{ q_p \}\right)\left|\psi\right\rangle_A\otimes\left|\tilde{\text{in}}\right\rangle_B,
\end{equation}
where
\begin{equation}
\mathcal{N}=N^{-M_{A}+1},
\end{equation}
and $M_A$ is the number of sites in lattice $A$ (or plaquette of $B$).

Under the action of the switch operator, a gauge invariant state in the $AB$ setting  becomes gauge fixed in the switched, $BA$ setting.  Moreover, the gauging operator, when restricted to a specific global charge sector, is unitary, since $W$ is an isometry (the proof is a stright-forward generalization to what we did in the one dimensional case). 
Choosing the local charges of $AB$ setting fixes the "out" state, it is $\left|\text{out}\right\rangle_A=\underset{p}{\bigotimes}\left|q_p\right\rangle$. Again, for the global charge sector $q=0$, where $\mathcal{Z}$ has eigenvalue $1$,  we can gauge with all $q_p=0$, and thus we get $\left|\text{out}\right\rangle = \underset{p}{\bigotimes}\left|0\right\rangle_p$, in accordance with (\ref{2ddualH}).

Here, too, for the sake of making the process feasible, just like in the one dimensional case we can begin with $U$ and then post-select the state $\left|\text{out}\right\rangle$ by projecting each $A$ site (or $B$ plaquette) separately to the $Z_p$ eigenstate $\left|0\right\rangle_p$.

\subsection{The two local symmetries of the $BA$ setting}

As expected, the dual Hamiltonian we found (\ref{2ddualH}) is that of a pure $\mathbb{Z}_N$ lattice gauge theory \cite{horn_hamiltonian_1979}. Not only this is not a case of self duality - the number of degrees of freedom is not the same; in the original, $A$ formulation, the degrees of freedom resided on the sites, while in the dual, $B$ formulation, they reside on the links. Yet, the local Hilbert spaces of both models are of the same dimension - $N$, so naively speaking, the Hilbert space dimension has increased significantly.

\begin{figure}[h]
    \centering
\includegraphics[width=0.2\textwidth]{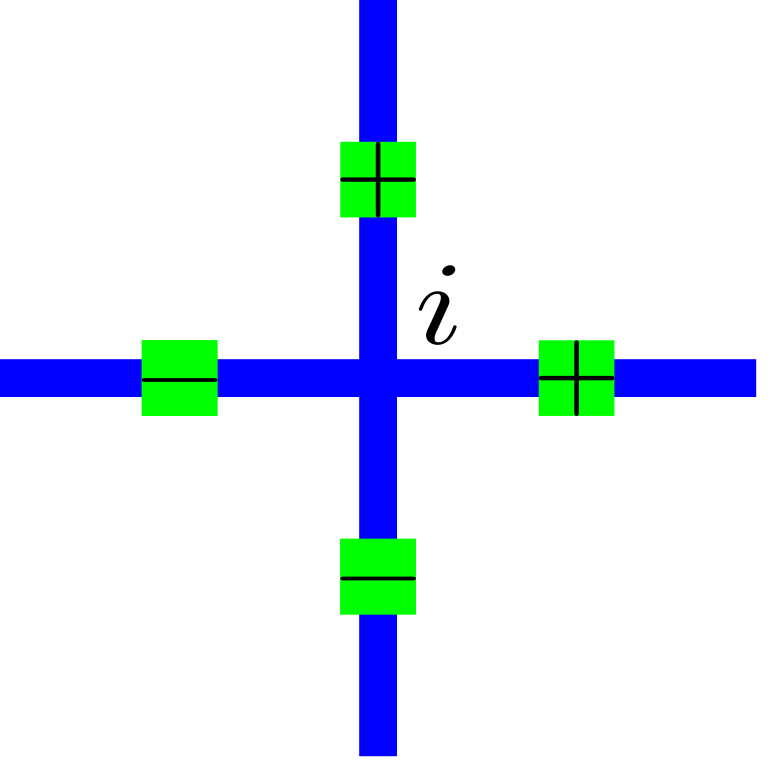}
    \caption{Notation of positive and negative links emanating from a site.}
    \label{fig:localsym}
\end{figure}

This issue of degrees of freedom is resolved by the additional symmetries of the dual theory; it is a gauge theory, and thus the Hamiltonian $\tilde{H}$ (\ref{2ddualH}), besides the global $\mathbb{Z}_N$ symmetry implemented by 
\begin{equation}\label{global2dZ}
    \tilde{\mathcal{Z}}=\prod_l \tilde{Z}_l,
\end{equation}
also has a local, gauge symmetry. This implemented by the unitaries
\begin{equation}\label{2d dual local symmetry}
    \tilde{G}_i=\underset{l\in i+}{\prod}\tilde{Z}_{l}\underset{l\in i-}{\prod} \tilde{Z}_{l}^\dagger
\end{equation}
for each $B$ site $i$, where $i+$ includes the links going out of $i$  and $i-$ stands for the ingoing ones, as shown in Fig.\ref{fig:localsym}.

While this is well known already, note that the $BA$ setting has \emph{two} types of gauge symmetries, the one of $B$ alone, as well as the auxiliary one which we artificially introduced for our construction. The first one is attached to the sites, and implemented by $\tilde{G}_i$ from (\ref{2d dual local symmetry}), while the second is associated with the links, and implemented by $\tilde{W}_l$ (\ref{tildeWl}). In the first, the gauge fields $B$ are really gauge fields, but in the other one, they play the role of matter degrees of freedom!

Not only
\begin{equation}
    \left[\tilde{G}_i,\tilde{G}_j\right]=\left[\tilde{G}_i,\tilde{H}'\right]=0,\quad \forall i,j
\end{equation}
and
\begin{equation}
    \left[\tilde{W}_l,\tilde{W}_m\right]=\left[\tilde{W}_l,\tilde{H}'\right]=0,\quad \forall l,m;
\end{equation}
also,
\begin{equation}
    \left[\tilde{W}_l,\tilde{G}_i\right]=0,\quad \forall l,i;
\end{equation}
implying that both symmetries can coexist. This is similar to what we had in the opposite direction, that the $AB$ gauge symmetry did not violate the $A$ global symmetry, and it implies that when we perform the procedure in the opposite direction, the $\tilde{W}_l$ static charges we gauge with must agree with the superselection sector dictated by the $G_i$ static charges, using the relations
\begin{equation}\label{globallocal}
    \tilde{\mathcal{Z}}=\underset{l}{\prod}\tilde{W}_l
\end{equation}
and
\begin{equation}\label{locallocal}
    \tilde{G}_i=\underset{l\in i+}{\prod}\tilde{W}_l \underset{l\in i-}{\prod}\tilde{W}_l^{\dagger}
\end{equation}
\subsection{Inverting the process}

When we invert the procedure and start from $B$, the $\mathbb{Z}_N$ lattice gauge theory, comes the question of which symmetry is the one that we lift to a local one when going to the $BA$ setting: the global symmetry implemented by $\tilde{\mathcal{Z}}$ (\ref{global2dZ})  or the local one, implemented by the $\tilde{G}_i$ operators (\ref{2d dual local symmetry}). The answer is that these are two valid ways to interpret our gauging; we want to find a transformation affecting $\tilde{X}_l$ on single links, and both the global and local transformations of $B$ transform several of those in parallel: either everywhere, or around a site. Therefore, and adding the fact that both symmetries coexist - we can say that we gauge both.

To minimally couple the Hamiltonian $\tilde{H}$ (\ref{2ddualH}), we introduce extra   $\mathbb{Z}_N$ Hilbert spaces on the plaquettes $p$. The gauged Hamiltonian is then $\tilde{H}'_{BA}$ (\ref{tildeHBA}), without dynamics for the $A$ degrees of freedom. We choose to begin with gauge fixing using $\left|\text{out}\right\rangle$ and connect the systems with $U^{\dagger}$, the inverse gauging operator defined in (\ref{U2d}).

\begin{figure}[h]
    \centering
    \includegraphics[width=0.47\textwidth]{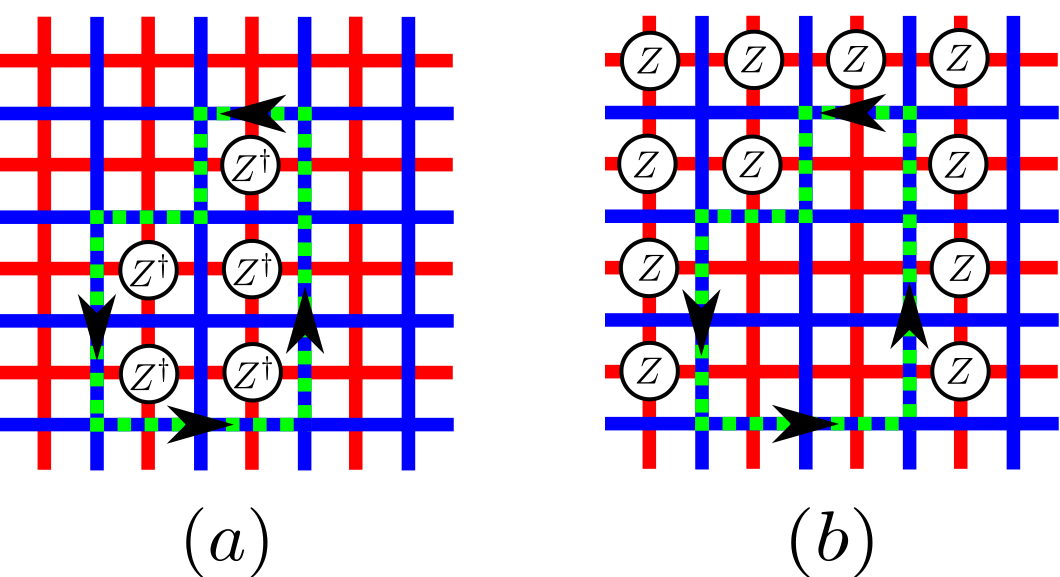}
    \caption{Gauguing in the inverse direction: we gauge closed oriented loops of $\tilde{X}$ (and $\tilde{X}^{\dagger}$ opeartors) by multiplying with $Z^{\dagger}_p$ or $Z_p$ on each plaquette within or out of the loop.}
    \label{fig:gauging inverse}
\end{figure}
When the gauging of other operators is considered, however, one should not only consider that of globally invariant ones, but also those invariant under the original local symmetry of $B$ (e.g. functions of plaquette interactions, such as $\tilde{X}_{p1}\tilde{X}_{p2}\tilde{X}^{\dagger}_{p3}\tilde{X}^{\dagger}_{p4}$. Consider, for example, an oriented product of $X_p$ (or $X^{\dagger}_p$) operators along some closed path. These will be gauged either by multiplication with all the $Z^{\dagger}_p$ / $Z_p$ within or out of the loop (see Fig. \ref{fig:gauging inverse}), but here too the choice does not matter, since start from the $\left|\text{out}\right\rangle$ states.

The gauging transformation, which replaces the gauge fixing by gauge invariance with respect to $\tilde{W}_l$ (\ref{tildeWl}), is defined as usual, with
\begin{equation}
\tilde{W}\left(\{\tilde{q}_l\}\right)=\underset{l}{\prod}\frac{\overset{N-1}{\underset{n=0}{\sum}}\left(\tilde{W_l}e^{-i\tilde{q}_l}\right)^n}{N},
\end{equation}
where $M_B$ is the number of links in system $B$. The static charges must agree with both the global sector $\tilde{q}$ ($\tilde{\mathcal{Z}}$) and the static charges of the original local symmetry, $\tilde{q}_i$ ($\tilde{G}_i$). For the first we demand that
$\underset{l}{\sum}\tilde{q}_l = \tilde{q}$ (modulo $N$) and for the second, $\underset{l \in i+}{\sum}\tilde{q}_l - \underset{l \in i-}{\sum}\tilde{q}_l =0$ (modulo $N$), for every $i$, using (\ref{globallocal}) and (\ref{locallocal}), respectively.

The gauged state, using the regular procedure, will take the form
\begin{equation}
    \left|\tilde{\Psi}\right\rangle_{BA}=\tilde{\mathcal{N}}^{-1/2}\tilde{W}\left(\{q_i\}\right)\left|\text{out}\right\rangle_A\otimes\left|\tilde{\psi}\right\rangle_B.
\end{equation}
The normalization constant is identical to the one we had in the other direction, i.e. $\tilde{\mathcal{N}}=\mathcal{N}=N^{1-M_A}$ (see appendix.\ref{appendix: normalization}).

 The other details of the gauging procedure in this direction are straight-forward adaptions of the other direction, and do not require special discussion.

\section{Compact QED ($U(1)$)  in $2+1d$}

The above arguments also generalize to $U(1)$ - a continuous Abelian group. For that, we considerr $2+1$d compact QED (\cite{kogut_introduction_1979}) - a lattice gauge theory with $U(1)$ as its gauge group, which is obtained at the $N \rightarrow \infty$ limit of $\mathbb{Z}_N$ lattice gauge theories \cite{horn_hamiltonian_1979}.

Starting from the lattice gauge theory side, if we use the terminology of the previous section, means that we go from $B$ to $A$, and thus the gauge theory operators and states will be denote here with a $\tilde{\cdot}$.

As in any other lattice gauge theory, the gauge fields reside on the lattice's links. The Hilbert space on each link $l$ is that of a particle on a ring, hosting a canonical pair of phase and $U(1)$ angular momentum operators, $\tilde{\phi}$ and $\tilde{L}$,
\begin{equation}\label{canonical}
\left[\tilde{L}_l,\tilde{\phi}_{l'}\right]=-i\delta_{l,l'}
.\end{equation}
The angular momentum operator, playing the role of the electric field, has an integer, non bounded spectrum ($L\left|m\right>=m\left|m\right>$). ${\phi}$ plays the role of a compact  vector potential.
The canonical relation (\ref{canonical}) implies that $e^{iL}$ shifts phase eigenstates:
\begin{equation}
    e^{i\alpha L}\left|\phi\right\rangle=\left|\phi-\alpha\right\rangle.
\end{equation}

The Hamiltonian of this model (mostly referred to as the Abelian version of the Kogut-Susskind Hamiltonian \cite{kogut_hamiltonian_1975}) takes the form (keeping the notations of section \ref{2DZN}):
\begin{equation}\label{2dU1H}
\begin{aligned}
    \tilde{H}_B=\lambda\underset{l}{\sum}\tilde{L}_l^2+\sum_p\left( e^{-i(\tilde{\phi}_{p_1}+\tilde{\phi}_{p_2}-\tilde{\phi}_{p_3}-\tilde{\phi}_{p_4})}+h.c.\right)
\end{aligned}.
    \end{equation}
The first term, corresponding to the electric energy, is a sum over the lattice's links. The second term, of the magnetic energy, is a sum over all the plaquettes of the lattice.

This Hamiltonian has a global $U(1)$ symmetry, implemented by the unitary operator
\begin{equation}
    \tilde{\mathcal{Z}}_\theta=\underset{l}{\prod}e^{i\theta \tilde{L}_l}, \quad \forall \theta \in \left[0,2\pi\right)
,\end{equation}
as well as a local  $U(1)$ symmetry too, implemented by the local unitaries
\begin{equation}
\tilde{\mathcal{U}}_i = e^{i\theta \tilde{\mathcal{G}}_i}, \quad \forall \theta \in \left[0,2\pi\right)
\end{equation}
for every lattice site $i$, with the local generators (Gauss law operators)
\begin{equation}
   \tilde{\mathcal{G}}_i = \underset{l\in i+}{\sum} \tilde{L}_l -  \underset{l\in i-}{\sum} \tilde{L}_l.
\end{equation}

As we did in the $\mathbb{Z}_N$ models, we proceed by introducing another local symmetry and and extra set of gauge fields, residing on the plaquettes, identified as the $A$ system. With respect to this symmetry, in the $BA$ setting the $B$ degrees of freedom (gauge fields in the $B$ setting) will play the role of matter fields. We minimally couple the Hamiltonian  (\ref{2dU1H}) such that the new plaquette ($A$) degrees of freedom, $\Theta_p$, which reside in similar Hilbert spaces, will compensate the transformation of a single $\phi$ variable. This is achieved by
\begin{equation}
  \tilde{H}'_{BA}=\lambda\underset{l}{\sum}\tilde{L}_l^2+\sum_p\left( e^{-i(\tilde{\phi}_{p1}+\tilde{\phi}_{p2}-\tilde{\phi}_{p3}-\tilde{\phi}_{p4})}e^{i\Theta_p} +h.c.  \right) 
,\end{equation}
where $e^{i\Theta_p}$ transforms as 
\begin{equation}
   e^{i\Theta}_p\to e^{i\left(\theta_{p_1}+\theta_{p_2}-\theta_{p_3}-\theta_{p_4}\right)} e^{i\Theta_p} 
\end{equation}

We denote the $A$ operators canonically conjugate to $\Theta$ by $L$; then, the generators of these gauge transformations are 
\begin{equation}
    \tilde{G}_l=\tilde{L}_l+ L_{l-}-L_{l+}
,\end{equation}
where the $l\pm$ notation is as in Fig. \ref{fig:notations}(a). The transformations are implemented by the unitary operators
\begin{equation}
    \tilde{W}_l = e^{i\theta_l \tilde{G}_l}.
\end{equation}

One may then choose the starting point of the $BA$ gauging, $\left|\text{out}\right\rangle_A$, to be the product of $L_p$ eigenstates with eigenvalue $0$ on all the plaquettes, making $L_p$ or $e^{iL_p}$ the operators responsible for gauge fixing in the $BA$ sense. The switch operator $U$ thus needs to satisfy
\begin{equation}
U^{\dagger} \tilde{G}_l U = \tilde{L}_l
\end{equation}
which is achieved by the local unitaries
\begin{equation}
    U_{l}=e^{-i\left(L_{l-}-L_{l+}\right)\tilde{\phi}_{l}}
\end{equation}
combining together to $U=\underset{l}{\prod}U_l$.

Using the switch operator, we obtain 
\begin{equation}
\begin{aligned}
    H'_{AB}=U^\dagger\tilde{H}'_{BA}U=\underset{l}{\sum}\left(\tilde{L}_l- L_{l-}+ L_{l+}\right)^2 
\end{aligned}.
\end{equation}

It is straightforward to see that (in the absence of static charges) one needs to post-select the state $\left|\tilde{\text{in}}\right\rangle_B$ which is an eigenstate of all $\tilde{L}_p$ operators with eigenvalue $0$, to arrive at the dual Hamiltonian
\begin{equation}
\begin{aligned}
    H_A&=
        \left\langle \tilde{\text{in}} \right|_B H'_{AB} \left| \tilde{\text{in}} \right\rangle_B
    \\&=\underset{l}{\sum}\left(L_{l+}-{L}_{l-}\right)^2+\lambda\underset{p}{\sum}\left( e^{i{\Theta}_p}+e^{-i\Theta_p}\right)
\end{aligned}
\end{equation}
as in, for example \cite{drell_quantum_1979,kaplan_gauss_2018,bender_gauge_2020}.

Finally, we need to introduce the gauging map in this case. Generalizing the above procedure, and in parallel lines to \cite{haegeman_gauging_2015,zohar_removing_2019}, we set it to take the form
\begin{equation}
    \left|\text{out}\right\rangle_A \otimes\left|\tilde{\psi}\right\rangle_B
    \to\left|\Psi_G\right>=\mathcal{N}^{-\frac{1}{2}}\int\mathcal{D}\theta e^{i\sum_l \tilde{G}_l\theta_l} \left|\psi\right\rangle_A\otimes\left|\text{in}\right>_B
,\end{equation}
where $\mathcal{D}\theta=\Pi_l\frac{ d\theta_l }{2\pi}$. The normalization factor in this case is   $\mathcal{N}=\left(\frac{1}{2\pi}\right)^{M_B-1}$, with $M_B$ - the number of links of the lattice $B$. This gauging operator assumes no static charges of any kind since they have no physical importance in most cases. Generalizations which include them are possible and straight-forward.

 \section{Summary}
  
  In this work, we have demonstrated for a variety of Abelian lattice models, how one could implement duality transformations physically, as maps between dual states exhibiting the same observable physics. This can be done with feasible quantum simulation techniques. We did this by explicitly constructing the duality maps, which involved only two-body operations and single-body measurements, and hence they are, in principle, within the reach of current quantum simulation technologies. Obviously, this is a very general statement, which will depend on the simulated model, the experimental platform and the simulation method; yet, the fact that the procedure can be factorized to such operations already forms an almost necessary condition for the feasibility of quantum operations on current day quantum technologies.
  
  This work, however, only serves as a proof of concept; possible extensions of it could be generalizations to other types of models, including a rather more theoretical classification of models which allow for this procedure and others which do not. While both gauging \cite{haegeman_gauging_2015} and the construction of unitary switch operators \cite{zohar_half_2017} are possible for arbitrary symmetry groups, their combination towards a feasible duality transformation is not guaranteed a-priori to any models with a duality transformation. This also depends on the degrees of freedom contained in the physical systems (for example, this work only considered first quantized physics, and the extension to Fock spaces, either bosonic or fermionic, is not trivial and very interesting). In such a study, looking for other gauging schemes, and more importantly, perhaps, non-unitary (yet isometric) switch operators, will most likely be required.

    \section*{Acknowledgements}
  This research was supported by the Israel Science Foundation (grant No. 523/20).
  
   \appendix

\section{Normalization}\label{appendix: normalization}
Here we show how to relate the two normalization factors ($AB$ and $BA$) in the $2+1d$ $\mathbb{Z}_N$ case. both in the $AB$ sense and the $BA$ one.

First, note the relations
\begin{equation}\label{ab isometry}
    U^\dagger\left|\text{out}\right\rangle_A\otimes\left|\tilde{\psi}\right\rangle_B=\mathcal{N}^{-1/2}W(\left(\{q_i\}\right)\left|\psi\right\rangle_A\otimes\left|\tilde{\text{in}}\right\rangle_B
\end{equation}
- demonstrating how the switch operator maps a gauge fixed state in one setting to a gauge invariant in the switched setting,  
as well as
\begin{equation}\label{ba isometry}
    U\left|\psi\right\rangle_A\otimes\left|\tilde{\text{in}}\right\rangle_B=\tilde{\mathcal{N}}^{-1/2}\tilde{W}(\left(\{q_i\}\right)\left|\text{out}\right\rangle_A\otimes\left|\tilde{\psi}\right\rangle_B
\end{equation}
in the other direction.

In the $AB$ setting, the normalization is given by
\begin{equation}
\mathcal{N}= 
\left\langle\psi\right|_A\otimes\left\langle\tilde{\text{in}}\right|_B W \left(\left\{q_i\right\}\right)
\left|\psi\right\rangle_A\otimes\left|\tilde{\text{in}}\right\rangle_B
;\end{equation}
where we have used $W^2=1$, as a projection operator. Plugging $U^\dagger U$ from the left of $W$ and using (\ref{ba isometry}) we obtain
\begin{equation}
\mathcal{N}= 
\tilde{\mathcal{N}}^{-1/2}\left\langle\text{out}\right|_A\otimes\left\langle\tilde{\psi}\right|_B  \tilde{W}UW \left|\psi\right\rangle_A\otimes\left|\tilde{\text{in}}\right\rangle_B  
.\end{equation}
Using (\ref{ab isometry}) we find
\begin{equation}
\mathcal{N}^{1/2}\tilde{\mathcal{N}}^{-1/2}\left\langle\text{out}\right|_A\otimes\left\langle\tilde{\psi}\right|_B  \tilde{W} \left|\text{out}\right\rangle_A\otimes\left|\tilde{\psi}\right\rangle_B  =\sqrt{\mathcal{N}\tilde{\mathcal{N}}}
,\end{equation}
implying that
\begin{equation}
    \mathcal{N}=\tilde{\mathcal{N}}
.\end{equation}

\bibliography{ref}
\end{document}